\documentclass[a4paper]{jpconf}
\usepackage{graphicx}

\usepackage[utf8]{inputenc}	
\usepackage[T1]{fontenc}	

\usepackage[colorinlistoftodos,prependcaption,textsize=tiny]{todonotes}
\usepackage[colorlinks=true, allcolors=blue]{hyperref}
\usepackage{lineno}

\newcommand{\kdk}{\mbox{KDK}}
\newcommand{\DAMA}{\mbox{DAMA}}

\newcommand{\potf}{\mbox{${}^{40}\mbox{K}$}}
\newcommand{\arf}{\mbox{${}^{40}\mbox{Ar}$}}

\newcommand{\mnf}{\mbox{${}^{54}\mbox{Mn}$}}
\newcommand{\crf}{\mbox{${}^{54}\mbox{Cr}$}}
\newcommand{\irf}{\mbox{${}^{55}\mbox{Fe}$}}
\newcommand{\zns}{\mbox{${}^{65}\mbox{Zn}$}}

\newcommand{\cms}{\mbox{$\mbox{cm}^{2}$}}
\newcommand{\cmc}{\mbox{$\mbox{cm}{}^{3}$}}

\begin{document}
\title{The \kdk\ (potassium decay) experiment}


\author{P.C.F.~Di Stefano${}^1$, N.~Brewer${}^2$,  A.~Fija{\l}kowska${}^2$$^{,}$${}^6$$^{,}$${}^8$, Z.~Gai${}^9$, K.C.~Goetz${}^3$, R.~Grzywacz${}^2$$^{,}$${}^3$$^{,}$${}^7$, D.~Hamm${}^3$, P.~Lechner${}^5$, Y.~Liu${}^2$, E.~Lukosi${}^3$, M.~Mancuso${}^4$,  C.~Melcher${}^3$, J.~Ninkovic${}^5$, F.~Petricca${}^4$, B.C.~Rasco${}^2$, C.~Rouleau${}^9$, K.P.~Rykaczewski${}^2$, P.~Squillari${}^1$, L.~Stand${}^3$, D.~Stracener${}^2$, M.~Stukel${}^1$, M.~Woli\'nska-Cichocka${}^2$$^{,}$${}^6$$^{,}$${}^7$ and I.~Yavin }
\address{${}^1$ Department of Physics, Engineering Physics and Astronomy, Queen's University, Kingston, K7L~3N6, Canada\\ ${}^2$ Physics Division, Oak Ridge National Laboratory, Oak Ridge, Tennessee 37831, USA\\ ${}^3$ Department of Physics and Astronomy, University of Tennessee, Knoxville, TN 37966, USA\\ ${}^4$ Max-Planck-Institut f\"ur Physik, F\"ohringer Ring 6, D-80805 M\"unchen, Germany\\ ${}^5$ MPG~Semiconductor~Laboratory, Munich, 81739, Germany\\ ${}^6$ Heavy Ion Laboratory, University of Warsaw, Warsaw PL-02-093, Poland\\ ${}^7$ Joint Institute for Nuclear Physics and Application, Oak Ridge, Tennessee 37831, USA\\ ${}^8$ Department of Physics and Astronomy, Rutgers University, New Brunswick, NJ 08903, USA\\ ${}^9$ Center for Nanophase Materials Sciences, Oak Ridge National Laboratory, Oak Ridge, Tennessee 37831, USA}

\ead{distefan@queensu.ca}

\begin{abstract}
Potassium-40 (\potf) is a background in many rare-event searches and may well play a role in interpreting results from the \DAMA\ dark-matter search.  The electron-capture decay of \potf\ to the ground state of \arf\ has never been measured and contributes an unknown amount of background.  The \kdk\ (potassium decay) collaboration will measure this branching ratio using a \potf\ source, an X-ray detector, and the Modular Total Absorption Spectrometer at Oak Ridge National Laboratory.
\end{abstract}

\section{Introduction}
Many lines of evidence suggest that most of the matter in the universe is dark, and possibly made of some new type of exotic particle~\cite{peebles_growth_2017-1} .  Experiments have been seeking dark-matter particles for close to three decades.  The \DAMA\ collaboration claims to have discovered them based on an annual modulation in the spectra observed in  NaI detectors~\cite{bernabei_dark_2013-2}, but this claim is controversial, as it is in strong tension with a number of experiments that use different techniques and target materials~\cite{noauthor_taup_nodate}.  In addition, \DAMA\ has not provided a background model for its claim, and disputes external attempts to do so~\cite{kudryavtsev_expected_2010-1,pradler_unverified_2013-1}.  It has recently been pointed out that an unmeasured decay of \potf\ could strongly constrain the dark-matter interpretation of the \DAMA\ claim~\cite{pradler_unverified_2013-1}.  \potf\ decays to \arf\ by electron capture~\cite{mougeot_recommended_nodate}, emitting X-rays and Auger electrons having energy close to 3~keV, within the signal region for dark matter.  Most of these decays (EC*), with a  branching ratio of $BR*=(10.55 \pm 0.11)$\%, are to an excited state of \arf, which returns to ground state by emitting a 1.46~MeV $\gamma$ thereby allowing one to tag the low-energy quanta.  However, direct decays to the ground state (EC) are expected as well, with no means to tag the low-energy emissions.  These direct decays are expected to be 50 times less frequent than those to the excited state.  Untagged low-energy emissions from \potf\ constrain the time-independent portion of the signal claimed by \DAMA, and therefore its modulation fraction, which itself must be consistent with astrophysical distributions.
The \kdk\ (potassium decay) collaboration proposes to measure this classical, but rare, branching ratio, which is  the only known unique third forbidden electron-capture transition.

\section{Experimental setup}
\begin{figure}[h]
\includegraphics[width=0.45\textwidth]{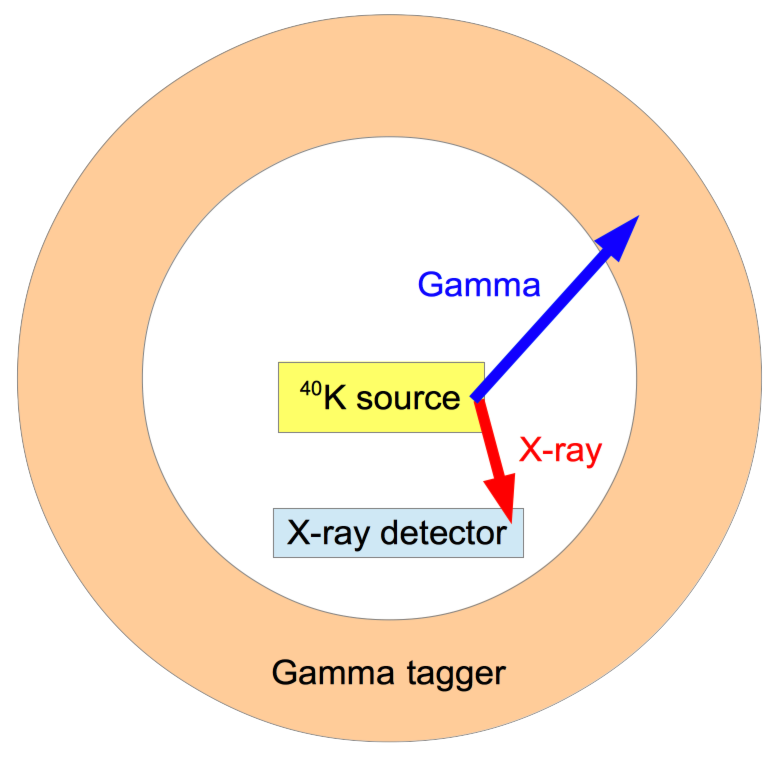}\hspace{2pc}%
\begin{minipage}[b]{14pc}\caption{\label{fig:setup} Setup of the \kdk\ experiment.}
\end{minipage}
\end{figure}

In \kdk, measuring $\zeta=BR/BR*$, the ratio of the EC and EC* branching ratios,  will involve the setup illustrated in Fig.~\ref{fig:setup}:
\begin{itemize}
\item a \potf\ source,
\item a small, sensitive detector to trigger on the X-rays from electron captures of all types (to ground state: EC, to excited state: EC*),
\item a large surrounding veto to tag 1.46~MeV $\gamma$s and distinguish between EC and EC* decays.
\end{itemize}
Our baseline design employs a source (discussed in Sec.~\ref{sec:Sources}) that is separate from the X-ray detector.

\begin{figure}[h]
\includegraphics[width=0.5\textwidth]{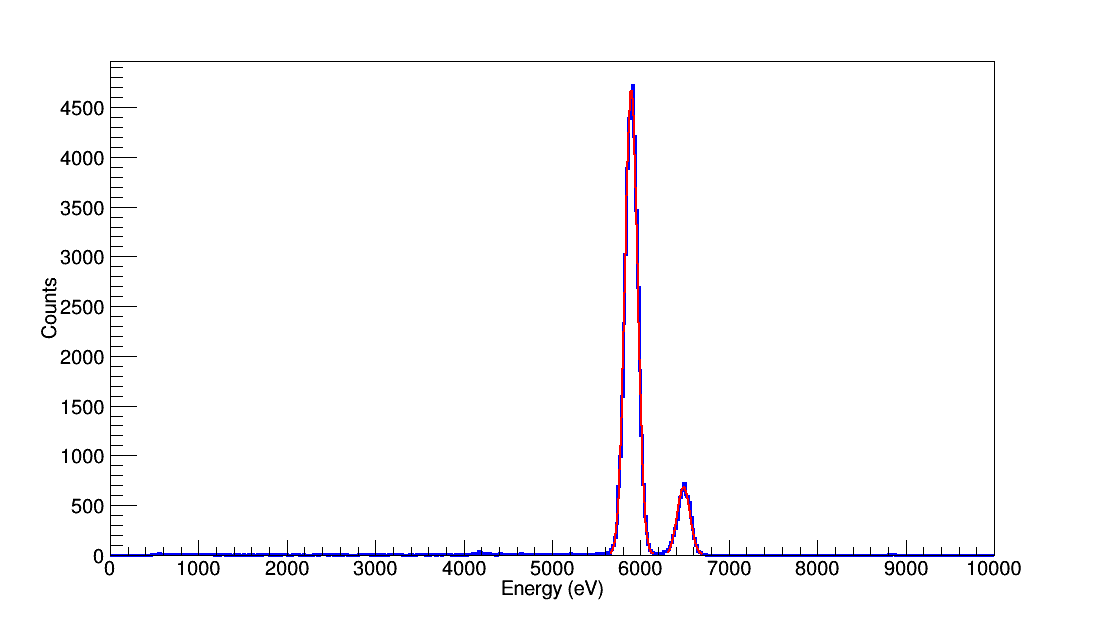}\hspace{2pc}
\begin{minipage}[b]{14pc}\caption{\label{fig:SDD} Spectrum of a \irf\ source obtained with a $\sim$~\cms\ SDD.  Resolution of the $K_{\alpha}$ and $K_{\beta}$ lines of Mn is 170~eV FWHM.}
\end{minipage}
\end{figure}

For the X-ray detector, the requirements are a threshold well below 3~keV, a surface area comparable to the $\sim 1 $~\cms\ of the source, and a limited amount of matter to reduce parasitic scattering of high-energy $\gamma$s.
Although off-the-shelf  large-area avalanche-photodiodes (APDs) of type S1315 from Radiation Monitoring Devices have been used initially (Fig.~\ref{fig:KDK}), they are being replaced by custom silicon drift-detectors with a better resolution of 170~eV~FWHM at 6~keV~\cite{niculae_optimized_2006-1} (Fig.\ref{fig:SDD}). 

The veto has stringent efficiency requirements.  Given that a single EC is expected for 50 EC*, to ensure that there is no more than one mis-tagged EC* for each true EC, the veto efficiency must be at least 98\%.  In other words, for each X-ray from EC* that triggers the X-ray detector, the detection efficiency of the related 1.46~MeV $\gamma$ must be at least 98\%, including geometry, absorption of the veto, and electronics.  This implies that the veto must cover nearly a full solid angle around the source and X-ray detector.  In addition, given that the attenuation length of 1.46~MeV $\gamma$s is relatively long (i.e. 5.7~cm in NaI~\cite{us_department_of_commerce_nist_nodate}), the veto must be quite thick (i.e. in NaI, to ensure that a $\gamma$ has a 99\% chance of interacting, the thickness must be at least 26~cm).  The Modular Total Absorption Spectrometer (MTAS)~\cite{wolinska-cichocka_modular_2014-1,rasco_nonlinear_2015-2,karny_modular_2016-1} at Oak Ridge satisfies these requirements. 
It is a $\sim 1$~Ton array of NaI(Tl) scintillators with a hole through the center where the \potf\ source and X-ray detector will be placed.
Its efficiency is of the correct order of magnitude for the purpose of \kdk, and will be determined precisely following the procedure in Sec.~\ref{sec:Sources}.  

Moreover, the collaboration has prepared a set of mechanical interfaces between the various subsystems.  Design constraints include:
\begin{itemize}
\item centring the source in MTAS, to maximize detection efficiency of $\gamma$s
\item minimizing distance between X-ray detector and source, to maximize detection efficiency of X-rays (to the first order, this ensures the experiment can be carried out with a reasonable exposure)
\item minimizing material near the source, to avoid parasitic scattering of $\gamma$ rays
\item cooling the X-ray detector to a temperature of the order of -20~C to reduce its noise
\item keeping the source and X-ray detector in a vacuum to allow the detector to be cooled without condensation
\end{itemize}
These constraints have been met using a thin-walled vacuum insert which houses the source and X-ray detector, and slides into the opening in MTAS.

Lastly, \kdk\ is also investigating an alternative to the separate source and X-ray detector.  KSr${}_2$I${}_5$ (KSI) is a recently-developed scintillator~\cite{stand_potassium_2013} with a density of 4.39~g/\cmc, and an excellent light yield of 95~photons/keV at 662~keV. 
This makes KSI an integrated source/detector solution whose advantages include high \potf\ activity (6.56~Bq/\cmc), and full absorption of all low-energy emissions from electron capture. System will be triggered on a  coincidence between two photomultipliers on the KSI, then a signal will be sought in MTAS.

\section{Efficiency and sources}
\label{sec:Sources}
\begin{figure}[h]
\includegraphics[width=0.5\textwidth]{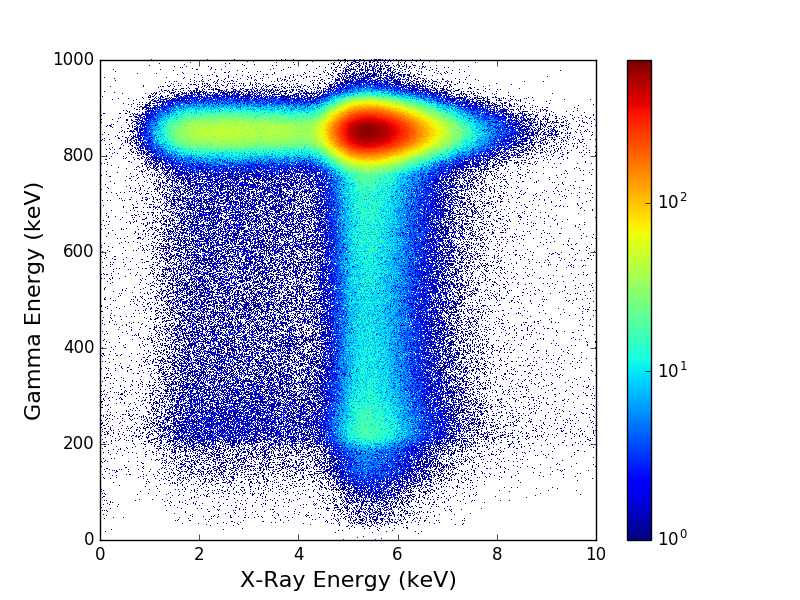}
\includegraphics[width=0.5\textwidth]{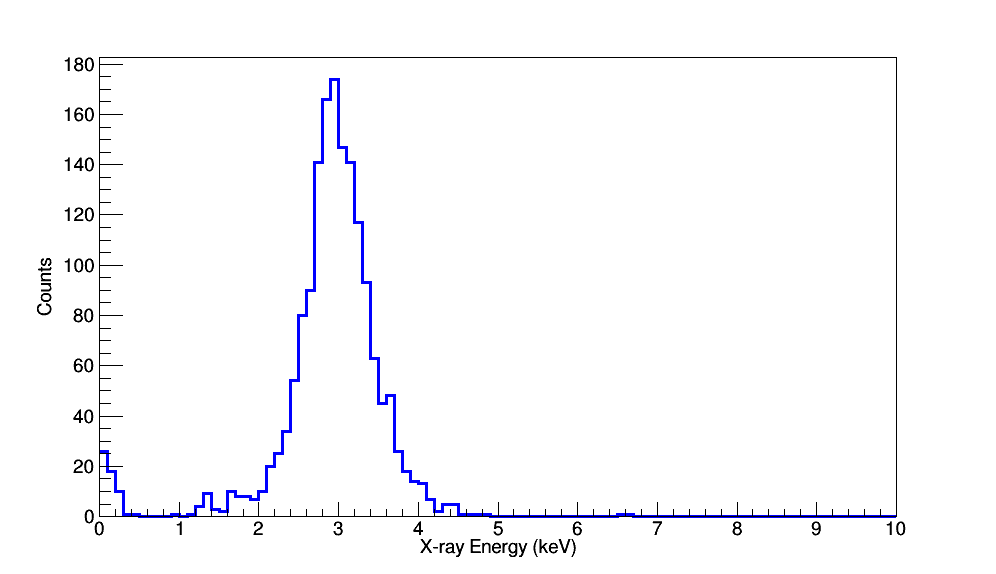}
\caption{\label{fig:KDK} Left: \mnf\ data ($\sim 10^6$ events) used to determine tagging efficiency of \kdk\ system, here using coincidences between 5~keV X-rays and 835~keV $\gamma$-rays.  Abscissas are energies in the APD X-ray detector, ordinates are those in the MTAS tagger.  Vertical smear represents partial energy deposits (eg Compton interactions) in MTAS.  Horizontal smear is attributed to superficial interactions in the APD (low energies) and to inhomogeneous response in plane of APD (high energies).  Right: APD spectrum of 3~keV X-rays in coincidence with 1.46~MeV $\gamma$s from our first \potf\ source.}
\end{figure}

To measure the ratio of branching ratios $\zeta=BR/BR*$ with a relative precision of 10\%, the efficiency of MTAS must be determined to a  precision of 0.2\%.  This will be achieved using radioactive isotopes with decay schemes similar to those of \potf, but with well-known branching ratios.  
The first such isotope is \mnf~\cite{mougeot_recommended_nodate}.  It decays by electron capture, emitting $\sim 5$~keV atomic radiation,  to an excited state of \crf, which relaxes by emitting an 835~keV $\gamma$, with an overwhelming branching ratio of $(99.9997 \pm 0.0003)$\%.  Data from a week-long \mnf\ calibration are presented in Fig.~\ref{fig:KDK}.
Another isotope, with a more complicated decay scheme, but higher energy $\gamma$s (1115~keV) is \zns~\cite{mougeot_recommended_nodate}.  Results from both isotopes are being analyzed.  Geant~4 Monte-Carlo simulations will be used to extrapolate the tagging efficiencies at 835~keV and 1115~keV to the 1460~keV relevant for \potf.

From the standpoint of the \potf\ source, the design goal is for it to contain $\sim 10^{18}$ atoms of \potf.  
The activity is equivalent to two bananas~\cite{hardisson_mineral_2001-1}.
The geometry will be a disk of area $\sim 1 $~\cms, and thickness of the order of $10 \ \mu$m to allow sufficient activity while limiting self-absorption of the 3~keV X-rays.
Two approaches have been followed.  In the first, we carried out ion-beam implantation of \potf\ on a thin carbon substrate.  
It was found that the number of \potf\ ions that could be implanted was very limited due to self-sputtering, preventing the source from reaching the required activity. 
 A more promising approach is to use thermal deposition of \potf-enriched KCl on a substrate. 
 Tests with natural KCl indicate a mass-transfer efficiency of 
 30--50\%. 
 We have demonstrated the production of the first \potf\ source obtained by thermal deposition of 3\% enriched KCl.  The source has about $3 \times 10^{17}$ \potf\ atoms in a  KCl film of about $9 \ \mu$m thickness.  Its 3~keV X-rays from EC* in coincidence with 1.46~MeV gammas have been observed, as illustrated in Fig.~\ref{fig:KDK}.  Work is proceeding to produce a stronger source starting from more-highly enriched KCl material.

Lastly, all the calibration and \potf\ sources are manufactured in the same geometry and on identical substrates to facilitate comparison.

\section{Blinding strategy}
To avoid introducing conscious or unconscious biases into our analysis, we have adopted a blinding strategy for \kdk.  It will be of "hidden signal box" type~\cite{klein_blind_2005-1}.  Here, the signal box represents  X-rays that are not coincident with $\gamma$s. All cuts will be developed on the coincident X-rays and $\gamma$s.  
The energy width of the coincident X-ray line will be used to set the width of the signal box.
Backgrounds in the signal region will be estimated from those adjacent to it.  Once these steps have been completed, the signal box will be opened and a likelihood analysis carried out on the data.
Should an error be found in the analysis after unblinding, we will report both the blinded result and the corrected one.

\section{Conclusions and outlook}
\kdk\ will measure the branching ratio of the electron capture decay of \potf\ to the ground state of \arf.  This result will help interpret the longstanding \DAMA\ claim for discovery of dark matter.  It will be of interest to the host of experiments about to test \DAMA\ with NaI detectors (including ANAIS~\cite{amare_anais_2016},
COSINE~\cite{adhikari_design_2017}, 
COSINUS~\cite{gutlein_cosinus_2017-1}, 
PICO-LON~\cite{fushimi_high_2016} and SABRE~\cite{froborg_sabre:_2016-1}), and rare-event searches in general since \potf\ is a common background. 
Moreover, \potf\ may radioactively heat the cores of planets, including Earth~\cite{murthy_experimental_2003-1}, and a precise measurement of the EC branching ratio will be valuable to geochronology and  \potf-\arf\ dating~\cite{nagler_pursuit_2000-1}. 

The main infrastructure of \kdk, including the MTAS tagger and interfaces, is in place.  First efficiency measurements have been performed using various radioactive sources and an APD as X-ray detector, and are being analyzed.  X-rays have been observed in coincidence with $\gamma$ rays from a first \potf\ source, while preserving the uncoincident X-rays as blind.  A stronger \potf\ source is being prepared.  
Moreover, new silicon drift-detectors are  being commissioned; they  offer an energy resolution improved by nearly an order of magnitude compared to the APDs,  insurance against possible backgrounds including $\beta^-$ decays of \potf.
Lastly, KSI scintillator may provide an additional tool for this measurement.

\ack
Paul Davis designed and supplied the APD electronics through the NSERC/University of Alberta MRS.  
Engineering support has also been contributed by Miles Constable and Fabrice R\'eti\`ere of TRIUMF, as well as by Koby Dering through the NSERC/Queen's MRS.
Work was performed at Oak Ridge National Laboratory, managed by UT-Battelle, LLC,
for the U.S.  Department of Energy under Contract DE-AC05-00OR22725.
Thermal deposition was conducted at the Center for Nanophase Materials Sciences, which is a DOE Office of Science User Facility.
Funding in Canada has been provided by NSERC through SAPIN and  SAP RTI grants.
Support has also been supplied by the Joint Institute for Nuclear Physics and Application.

\section*{References}
\bibliographystyle{my-iopart-num}
\bibliography{My_Library}

\providecommand{\newblock}{}
\begin{thebibliography}{10}
\expandafter\ifx\csname url\endcsname\relax
  \def\url#1{{\tt #1}}\fi
\expandafter\ifx\csname urlprefix\endcsname\relax\def\urlprefix{URL }\fi
\providecommand{\eprint}[2][]{\url{#2}}

\bibitem{peebles_growth_2017-1}
Peebles P~J~E 2017 {\em Nat. Astron.\/} {\bf 1} s41550--017--0057--017

\bibitem{bernabei_dark_2013-2}
Bernabei R {\em et~al.\/} 2013 {\em Int. J. Mod. Phys. A\/} {\bf 28} 1330022

\bibitem{noauthor_taup_nodate}
{TAUP} 2017 - {XV} {International} {Conference} on {Topics} in {Astroparticle}
  and {Underground} {Physics} {\textbar} {TAUP} 2017
  \urlprefix\url{https://taup2017.snolab.ca/}

\bibitem{kudryavtsev_expected_2010-1}
Kudryavtsev V, Robinson M and Spooner N 2010 {\em Astropart. Phys.\/} {\bf 33}
  91--96

\bibitem{pradler_unverified_2013-1}
Pradler J, Singh B and Yavin I 2013 {\em Phys. Lett. B\/} {\bf 720} 399--404

\bibitem{mougeot_recommended_nodate}
Mougeot X and Helmer R~G Recommended data
  \urlprefix\url{http://www.nucleide.org/DDEP_WG/DDEPdata.htm}

\bibitem{niculae_optimized_2006-1}
Niculae A {\em et~al.\/} 2006 {\em Nucl. Instr. Meth. Phys. Res. A\/} {\bf 568}
  336--342

\bibitem{us_department_of_commerce_nist_nodate}
US~Department~of Commerce N {NIST} {XCOM}: {Photon} {Cross} {Sections}
  {Database} \urlprefix\url{http://www.nist.gov/pml/data/xcom/index.cfm}

\bibitem{wolinska-cichocka_modular_2014-1}
Wolińska-Cichocka M {\em et~al.\/} 2014 {\em Nucl. Data Sheets\/} {\bf 120}
  22--25

\bibitem{rasco_nonlinear_2015-2}
Rasco B {\em et~al.\/} 2015 {\em Nucl. Instr. Meth. Phys. Res. A\/} {\bf 788}
  137--145

\bibitem{karny_modular_2016-1}
Karny M {\em et~al.\/} 2016 {\em Nucl. Instr. Meth. Phys. Res. A\/} {\bf 836}
  83--90

\bibitem{stand_potassium_2013}
Stand L {\em et~al.\/} 2013 Potassium strontium iodide: {A} new high light
  yield scintillator with 2.4\% energy resolution {\em 2013 {IEEE} {Nuclear}
  {Science} {Symposium} and {Medical} {Imaging} {Conference} (2013
  {NSS}/{MIC})\/} pp 1--3

\bibitem{hardisson_mineral_2001-1}
Hardisson A {\em et~al.\/} 2001 {\em Food Chem.\/} {\bf 73} 153--161

\bibitem{klein_blind_2005-1}
Klein J~R and Roodman A 2005 {\em Ann. Rev. Nucl. Part. Sci.\/} {\bf 55}
  141--163

\bibitem{amare_anais_2016}
Amaré J {\em et~al.\/} 2016 {\em arXiv:1601.01184 [astro-ph,
  physics:physics]\/}

\bibitem{adhikari_design_2017}
Adhikari G {\em et~al.\/} 2017 {\em arXiv:1710.05299 [astro-ph, physics:hep-ex,
  physics:physics]\/}

\bibitem{gutlein_cosinus_2017-1}
Gütlein A {\em et~al.\/} 2017 {\em Nucl. Instr. Meth. Phys. Res. A\/} {\bf
  845} 359--362

\bibitem{fushimi_high_2016}
Fushimi K~I {\em et~al.\/} 2016 {\em arXiv:1605.04999 [astro-ph,
  physics:physics]\/}

\bibitem{froborg_sabre:_2016-1}
Froborg F {\em et~al.\/} 2016 {\em arXiv:1601.05307 [hep-ex,
  physics:physics]\/}

\bibitem{murthy_experimental_2003-1}
Murthy V~R, Van~Westrenen W and Fei Y 2003 {\em Nat.\/} {\bf 423} 163--165

\bibitem{nagler_pursuit_2000-1}
Nägler T~F and Villa I~M 2000 {\em Chem. Geol.\/} {\bf 169} 5--16

\end{thebibliography}


\end{document}